\documentclass[aps,twocolumn,preprintnumbers,showpacs,showkeys]{revtex4-1}
\usepackage{epsfig}
\usepackage{amssymb,amsmath,amsfonts,amsthm,graphicx}
\graphicspath{{./Figures/}}                 
\newcommand{\beq}{\begin{equation}}
\newcommand{\eeq}{\end{equation}}
\newcommand{\bea}{\begin{array}}
\newcommand{\eea}{\end{array}}
\newcommand{\beqa}{\begin{eqnarray}}
\newcommand{\eeqa}{\end{eqnarray}}

\def\beqa{\begin{eqnarray}}
\def\eeqa{\end{eqnarray}}

\begin{document}

\title{New approach to canonical partition functions computation in $N_f=2$ lattice QCD  at finite baryon density }

\author{V.~G.~Bornyakov}
\affiliation{Institute for High Energy Physics NRC Kurchatov Institute,
142281 Protvino, Russia, \\
Institute of Theoretical and Experimental Physics NRC Kurchatov Institute, 117218 Moscow,
Russia \\
School of Biomedicine, Far Eastern Federal University, 690950 Vladivostok,
Russia}

\author{D.~L. Boyda}
\affiliation{School of Natural Sciences, Far Eastern Federal University, 690950 Vladivostok, Russia,\\
School of Biomedicine, Far Eastern Federal University, 690950 Vladivostok, Russia,\\
Institute of Theoretical and Experimental Physics NRC Kurchatov Institute, 117218 Moscow, Russia}

\author{V.~A. Goy}
\affiliation{School of Natural Sciences, Far Eastern Federal University, 690950 Vladivostok, Russia,\\
School of Biomedicine, Far Eastern Federal University, 690950 Vladivostok, Russia,\\
Institute of Theoretical and Experimental Physics NRC Kurchatov Institute, 117218 Moscow, Russia}

\author{A.~V.~Molochkov}
\affiliation{School of Biomedicine, Far Eastern Federal University, 690950 Vladivostok, Russia,\\
Institute of Theoretical and Experimental Physics NRC Kurchatov Institute, 117218 Moscow, Russia}

\author{Atsushi Nakamura}
\affiliation{School of Biomedicine, Far Eastern Federal University, 690950 Vladivostok, Russia,\\
Research Center for Nuclear Physics (RCNP), Osaka University, Ibaraki, Osaka, 567-0047, Japan,\\
Theoretical Research Division, Nishina Center, RIKEN, Wako 351-0198, Japan
}

\author{A.~A. Nikolaev }
\affiliation{School of Biomedicine, Far Eastern Federal University, 690950 Vladivostok, Russia,\\
Institute of Theoretical and Experimental Physics NRC Kurchatov Institute, 117259 Moscow, Russia}

\author{V.~I. Zakharov}
\affiliation{Institute of Theoretical and Experimental Physics NRC Kurchatov Institute, 117259 Moscow,
Russia \\
School of Biomedicine, Far Eastern Federal University, 690950 Vladivostok, Russia \\
Moscow Institute of  Physycs and Technology, Dolgoprudny, Moscow Region, 141700 Russia,
}

\date{2016}
\date{\today}
\begin{abstract}
 We propose and test a new approach to computation  of canonical partition functions in lattice QCD at finite density.
 We suggest a few steps procedure. We first compute numerically the quark number density for imaginary chemical potential $i\mu_{qI}$. Then we restore the grand canonical partition function for imaginary chemical potential using fitting procedure for the quark number density. Finally we compute the canonical partition functions using high precision numerical Fourier transformation. Additionally we compute the canonical partition functions using  known method of the hopping parameter expansion and compare results obtained by two methods in the deconfining as well as in the confining phases.
 The agreement between two methods indicates the validity of the new method. Our numerical results are obtained in two flavor lattice QCD with clover improved Wilson fermions.
\end{abstract}



\maketitle

\section{Introduction}
\label{sec:introduction}

One of the most important research targets in high energy and nuclear physics
is to reveal the hadronic matter phase structure
at finite temperature and density.
 Experiments at most important modern accelerators  RHIC (BNL) \cite{Adams:2005dq},  LHC (CERN) \cite{Aamodt:2008zz}
  and future experiments
 FAIR (GSI) and NICA (JINR)  are devoted to such studies.
Towards this goal, experimental, observational and theoretical efforts
have been made.
The lattice QCD numerical simulations have a mission to provide data from the first principle
calculations.
Indeed, at finite temperature with zero chemical potential,
the phase structure was satisfactorily investigated.
But it is very difficult to study the finite density regions by the lattice QCD
because of the infamous ``sign problem'':
The fermion determinant at nonzero baryon chemical potential $\mu_B$, $\det\Delta(\mu_B)$, is in general not real.
This makes impossible to apply standard Monte Carlo techniques to computations with the partition function
\beq
Z_{GC}(\mu_q,T,V) = \int \mathcal{D}U (\det\Delta(\mu_q))^{N_f} e^{-S_G},
\label{Eq:PathIntegral}
\eeq
where $S_G$ is a gauge field action, $\mu_q=\mu_B/3$ is quark chemical potential,   $T=1/(aN_t)$ is temperature, $V=(aN_s)^3$ is volume, $a$ is lattice spacing, $N_t, N_s$ - number of lattice sites in time and space directions.
There have been many trials~\cite{Muroya:2003qs,Philipsen:2005mj,deForcrand:2010ys} and yet it is still
very hard to get reliable results at $\mu_B/T>1$,
 see Ref.~\cite{Nagata:2012pc}.
In this paper, we discuss another trial, ``Canonical approach'',
which may boost us to go beyond current standard methods.
The canonical approach was studied in a number of papers \cite{deForcrand:2006ec,Ejiri:2008xt,Li:2010qf,Li:2011ee,Danzer:2012vw,Gattringer:2014hra,Fukuda:2015mva,Nakamura:2015jra}
We suggest new method to compute canonical partition function $Z_C(n, T, V)$, which allows to compute $Z_C(n, T, V)$ for large values of $n$ where $n$ is a net number of quarks and antiquarks. We will show that results
for $Z_C(n, T, V)$ obtained with the new method are in a good agreement with results obtained with the known method of hopping parameter expansion.

The new method is based on simulations at the imaginary
chemical potential. We explain the details of the new method in the next section. In section \ref{simulation} details of numerical simulations
including explanation of main features of the hopping parameter expansion are presented.
Numerical results for few values of temperature in both confinement and deconfinement phases and comparison with the hopping parameter expansion are presented in section \ref{mainresults}.
Finally, we formulate our conclusions in section \ref{conclusions}.

\section{New approach to computation of canonical partition function}
\label{sec:approach}

The canonical approach is based on the following relations. First, this is a relation between
grand canonical partition function $Z_{GC}(\mu_q, T, V)$ and the canonical one $Z_C(n, T, V)$

\begin{eqnarray}
Z_{GC}(\mu, T, V)=\sum_{n=-\infty}^\infty Z_C(n,T,V)\xi^n,
\quad
\label{ZG}	
\end{eqnarray}
where
$\xi=e^{\mu_q/T}$ is the fugacity and eq.~(\ref{ZG})
is called fugacity expansion.
The inverse of this equation can be presented in the following form
\cite{Hasenfratz:1991ax}
\begin{eqnarray}
Z_C\left(n,T,V\right)=\int_0^{2\pi}\frac{d\theta}{2\pi}
e^{-in\theta}Z_{GC}(\theta,T,V).
\label{Fourier}
\end{eqnarray}
In the right hand side of eq.~(\ref{Fourier}) we  see the grand canonical partition function  $Z_{GC}(\theta,T,V)$
for imaginary chemical potential $\mu_q=i\mu_{qI} \equiv iT\theta$ . It is known that standard Monte Carlo simulations are
possible for this  partition function since fermionic determinant is real for imaginary $\mu_q$.

\begin{figure}[htb]
\centering
\includegraphics[width=0.35\textwidth,angle=0]{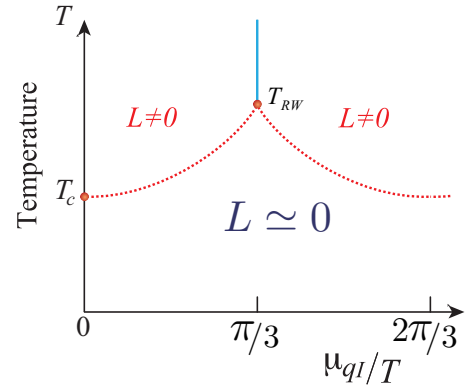}%
\vspace{0cm}
\caption{Schematical figure of Roberge-Weiss phase structure in the pure imaginary chemical
potential regions.
$T_c$ is the confinement/deconfinement crossover point at zero chemical potential,
$L$ is the Polyakov loop.
The vertical line $(T \ge T_{RW},\mu_I/T=\pi/3)$ shows the first order phase transition.
The dashed line is a crossover which can change to the first order phase transition
for large or small quark masses.
}
\label{RW_ph_d}
\end{figure}

The QCD partition function  $Z_{GC}$ is a periodic function of $\theta$: $Z_{GC}(\theta) = Z_{GC}(\theta+2\pi/3)$.
As a consequence of this periodicity the canonical partition functions $Z_C(n,T,V)$  are nonzero only for $n=3k$.
This symmetry is called Roberge-Weiss symmetry \cite{Roberge:1986mm}.
 QCD possesses a rich phase structure at nonzero $\theta$, which depends on the number of
flavors $N_f$ and the quark mass $m$. This phase structure is shown in
Fig.~\ref{RW_ph_d}.  $T_c$ is the confinement/deconfinement crossover point at
zero chemical potential.  The line $(T \ge T_{RW},\mu_I/T=\pi/3)$ indicates the
first order phase transition.  On the curve between $T_c$
and $T_{RW}$, the transition is expected to change from the crossover to the first order
for small and large quark masses, see e.g. \cite{Bonati:2014kpa}.

 Quark number density $n_q$ for $N_f$ degenerate quark flavours is defined by the following equation:
\beq
\frac{n_{q}}{T^{3}} = \frac{1}{VT^{2}}\frac{\partial}{\partial \mu_q}\ln
Z_{GC} \label{eq:n_Z}   \\
\eeq
\beq
=\frac{N_{f}N_{t}^{3}}{N_s^3 Z_{GC}} \int \mathcal{D}U e^{-S_G} (\det\Delta(\mu_q))^{N_f}
\mathrm{tr}\left[\Delta^{-1}\frac{\partial \Delta}{\partial \mu_q/T}\right] \nonumber
\label{density1}
\eeq
It can be computed numerically for imaginary chemical potential. Note, that for the imaginary chemical potential $n_q$ is also purely imaginary: $n_q = i n_{qI}$.

From eqs.~(\ref{ZG}) and (\ref{density1}) it follows that densities $n_{q}$ and $n_{qI}$ are related  to $Z_C(n,T,V)$ (below we will use the notation $Z_n$ for the ratio
$Z_C(n,T,V) / Z_C(0,T,V)$)  by equations
\beq
\label{density_re}
n_{q}/T^3  = {\cal{N}}\frac{2\sum_{n>0} n Z_n \sinh(n\theta)}{1+2\sum_{n>0} Z_n \cosh(n\theta)}\,,
\eeq
\beq
\label{density2}
n_{qI}/T^3  = {\cal{N}}\frac{2\sum_{n>0} n Z_n \sin(n\theta)}{1+2\sum_{n>0} Z_n \cos(n\theta)}\,,
\eeq
where ${\cal{N}}$ is a normalization constant, ${\cal{N}}=\frac{N_t^3 }{N_s^3}$.
Our suggestion is to compute $Z_n$ using equation (\ref{density2}).

One way to do this is to fit numerical data for
$ n_{qI}$ to this functional form with finite number of terms in both numerator and denominator with $Z_n$ treated
as a fitting parameters. We tried this method and found that it is difficult to obtain reliable values of $Z_n$.

We found more successful the following approach.
One can compute $Z_{GC}(\theta,T,V)$ using numerical data for $n_{qI}/T^3$ via numerical integration over imaginary chemical potential:
\beq
L_Z(\theta) \equiv \log\frac{Z_{GC}(\theta,T,V)}{Z_{GC}(0,T,V)}  = - V \int_{0}^{\theta} d \tilde{\theta}~n_{qI}(\tilde{\theta})\,,
\label{integration_1}
\eeq
where we omitted $T$ and $V$ from the grand canonical partition function
notation.
Then $Z_n$ can be computed as
\beq	
Z_n = \frac{\int_0^{2\pi}\frac{d\theta}{2\pi} e^{-in\theta} e^{L_Z(\theta)} }{ \int_0^{2\pi}\frac{d\theta}{2\pi}
 e^{L_Z(\theta)} }
\label{Fourier_2}
\eeq

In the present work we use modified version of this approach.
Instead of numerical integration in (\ref{integration_1})
we fitted $n_{qI}/T^3$ to theoretically motivated functions
of $\mu_{qI}$.

It is known that the density of noninteracting  quark gas is described  by
\beq
n_q/T^3 = N_f \Bigl ( 2\frac{\mu_q}{T}  + \frac{2}{\pi^2} \Bigl (\frac{\mu_q}{T} \Bigr )^3 \Bigr ).
\eeq
This allows to assume that in the deconfinement phase the density
can be fitted to the polynomial function.
Indeed, it was shown in Ref.~\cite{Takahashi:2014rta},  where the same lattice action was simulated,
that such function describes the number density in the deconfinement phase quite well.
This observation was also confirmed in Ref.~\cite{Gunther:2016vcp} where $N_f=2+1$ lattice QCD with physical quark masses was studied.
We thus fit the data for $n_{qI}$ to an odd power polynomial of $\theta$
\beq
n_{qI}(\theta)/T^3 = \sum_{n=1}^{n_{max}} a_{2n-1} \theta^{2n-1}\,,
\label{eq_fit_polyn}
\eeq
in the deconfining phase.

It is well known that in the confining phase (below $T_c$) the hadron resonance gas model provides
good description of the chemical potential dependence of thermodynamic observables
\cite{Karsch:2003zq}.
Thus it is reasonable to fit the density to a Fourier expansion
\beq
n_{qI}(\theta)/T^3 = \sum_{n=1}^{n_{max}} f_{3n} \sin(3n \theta)
\label{eq_fit_fourier}
\eeq
Again this type of the fit was  used in Ref.~\cite{Takahashi:2014rta} and conclusion was made that it works well.

 Using these fits and eqs.~(\ref{integration_1}), (\ref{Fourier_2})  we obtained very promising results for the canonical partition functions $Z_n$ as it will be shown in the section \ref{mainresults}.

\section{Simulation settings}
\label{simulation}
To demonstrate our method we make simulations of the lattice QCD with $N_f=2$ clover improved Wilson quarks and Iwasaki improved gauge field action
\begin{eqnarray}
  S   &=& S_g + S_q, \\
  S_g &=&
  -{\beta}\sum_{x,\mu\nu}  \left(
   c_0 W_{\mu\nu}^{1\times1}(x)
   + c_1 W_{\mu,\nu}^{1\times2}(x) \right), \\
  S_q &=& \sum_{f=u,d}\sum_{x,y}\bar{\psi}_x^f \Delta_{x,y}\psi_y^f,
  \label{eq:action}
\end{eqnarray}
where $\beta=6/g^2$, $c_1=-0.331$, $c_0=1-8c_1$,
\begin{eqnarray}
 \Delta_{x,y} &=& \delta_{xy} -\kappa\sum_{i=1}^3 \{(1-\gamma_{i})U_{x,i}\delta_{x+\hat{i},y} \nonumber \\&&
    +(1+\gamma_{i})U_{y,i}^{\dagger}\delta_{x,y+\hat{i}}\}
       \nonumber \\ &&
   -{\kappa} \{e^{a\mu_q}(1-\gamma_{4})U_{x,4}\delta_{x+\hat{4},y}   \nonumber\\ &&
    +e^{-a\mu_q}(1+\gamma_{4})U_{y,4}^{\dagger}\delta_{x,y+\hat{4}}\}   \nonumber \\ &&
   -\delta_{xy}{c_{SW}}{\kappa}\sum_{\mu<\nu}\sigma_{\mu\nu}P_{\mu\nu},
\label{eq:fermact}
\end{eqnarray}
$P_{\mu\nu}$  is the clover definition of the lattice field strength tensor,
$c_{SW} =(W^{1\times 1})^{-3/4}=(1-0.8412\beta^{-1})^{-3/4}$ is the Sheikholeslami–Wohlert coefficient.

We simulate $16^3 \times 4$ lattices at temperatures $T/T_c=1.35, 1.20$ and 1.08 in the deconfinemnt phase  and $0.99, 0.93, 0.84$ in the confinement phase along the line of constant physics with
 $m_{\pi}/m_{\rho}=0.8$. All parameters of the action, including $c_{SW}$ value were borrowed from the WHOT-QCD collaboration paper~\cite{Ejiri:2009hq}. We compute the number density on samples of $N_{conf}$ configurations with $N_{conf}=1800$, using every 10-th trajectory produced with Hybrid Monte Carlo algorithm.

We employ the hopping parameter expansion to compute $Z_n$ and compare with $Z_n$ values obtained with our new method.
Below we describe  the hopping parameter expansion.
The Wilson Dirac operator from (\ref{Eq:PathIntegral}) may be written in the form
\beq
\label{eq:D_op_Q}
\Delta = I - \kappa Q\,,
\eeq
both in the case of standard and clover improved Wilson fermions. Then one can rewrite the fermionic determinant in the following way~\cite{Nakamura:2015jra}:
\beq
\label{eq:Det_expansion}
\det\Delta = exp\Bigl[ \mathrm{Tr}\,ln \left( I - \kappa Q \right) \Bigr] = exp\Bigl[ - \sum_{n = 1}^{\infty} \frac{\kappa^n}{n} \mathrm{Tr} Q^n \Bigr]\,.
\eeq
The chemical potential is introduced in the Dirac operator in the form of $e^{\pm a\mu_q}$ multipliers in the temporal direction links
$U_{x,4}$. The expansion in (\ref{eq:Det_expansion}) is in fact  expansion over the closed paths on the lattice, and thus  (\ref{eq:Det_expansion}) can be rewritten as
\beq
\label{eq:Det_expansion_W_n}
\det\Delta = exp\Bigl[ \sum_{n = - \infty}^{\infty} W_n \xi^n \Bigr]\,,
\eeq
where $n$ is number of windings in the temporal direction, $W_n$ are complex coefficients which are called winding numbers. They satisfy the property $W_{-n} = W_n^*$.
In the case of the imaginary chemical potential the expansion (\ref{eq:Det_expansion_W_n}) will become
\beq
\label{eq:Det_expansion_im_muq}
\det\Delta = e^{W_0} e^{2\sum_{n = 1}^{\infty} \Bigl( \text{Re}[W_n]\cos (n \theta) - \text{Im}[W_n]\sin (n \theta) \Bigr) }\,.
\eeq

It is important to note that hopping parameter expansion (\ref{eq:Det_expansion}) converges properly only for heavy quark masses \cite{Nakamura:2015jra}. Our simulations were performed for the quark masses from this range.

We compute the traces in (\ref{eq:Det_expansion})
and in computation of the number density
using stochastic estimator method. $N_{stoch}=1000$  stochastic vectors are used to compute the traces in (\ref{eq:Det_expansion})  in the deconfinement phase and 200 stochastic vectors in the confinement phase. For the number density computations we take $N_{stoch}=600$. We have checked that further increasing of $N_{stoch}$ does not change our results.

\section{Fits of the number density to eqs.~(\ref{eq_fit_polyn}), (\ref{eq_fit_fourier})  }
\label{mainresults}

\begin{table*}[ht]
\begin{tabular}{|c|c|c|c|c|c|c|} \hline
$T/T_c$  & $a_1$ & $ a_3$& $ a_5$ &$\chi^2/N_{dof}, N_{dof}$  & $ 2c_2$ & $  24c_4$ \\  \hline
1.35(7)       & 4.671(2)&-0.991(4)& -               & 0.67, 24 &  4.682(11)    & 5.8(5) \\
1.20(6)       &4.409(6) &-1.032(31)& -0.165(32)& 0.70, 16&  4.403(14) & 7.8(9) \\
1.08(5)       &3.880(17)&-1.62(21)& -0.59(0.47)& 1.10, 9&  3.877(19)  & 8.0(10) \\ \hline
\end{tabular}
\caption{Results of fitting data for $n_{qI}/T^3$ in the deconfinement phase to function (\ref{eq_fit_polyn}). }
\label{table_results_1}
\end{table*}

In Ref.~\cite{Takahashi:2014rta} it was shown that the number density can be well described by a polynomial of $\theta$ in the deconfining phase (above $T_{RW}$) and by a few terms of the Fourier expansion below $T_c$. We will use this analysis and improve it in a few directions. Firstly, we collected higher statistics and reduced the statistical errors substantially in comparison with Ref.~\cite{Takahashi:2014rta}. Secondly, we simulate larger lattices: $16^3$ in spatial directions instead of $16^2 \times 8$ in Ref.~\cite{Takahashi:2014rta}. As a result we come to more solid statements about fitting.

\begin{figure}[htb]
\centering
\includegraphics[width=0.35\textwidth,angle=270]{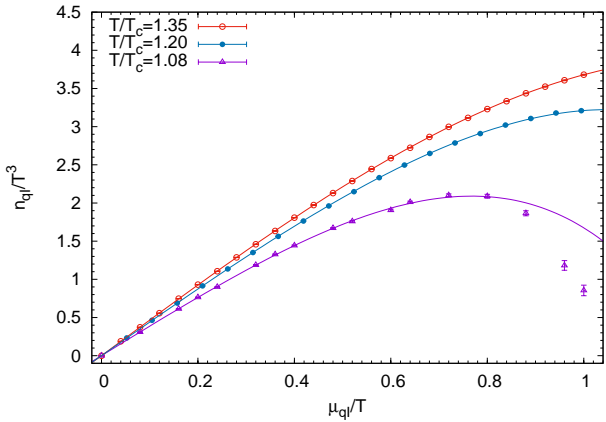}%
\vspace{0cm}
\caption{Imaginary density as function of $\theta$ in the deconfinement phase at temperatures $T/T_c=1.35, 1.20, 1.08$.
The curves show fits to function (\ref{eq_fit_polyn}). }
\label{density_deconf}
\end{figure}

In Fig.~\ref{density_deconf} we show results for $n_{qI}$ as a function of $\theta$ for $T/T_c=1.35, 1.20, 1.08$ for the range
of $\theta$ values between 0.0 and $\pi/3$. $n_{qI}$ is a continuous function over this range of $\theta$.
Results of the fits to function (\ref{eq_fit_polyn})
are presented in Table~\ref{table_results_1} and also shown in the figure. For $T/T_c=1.35$ we obtained
very good fit with $n_{max}=2$
with  $\chi^2/N_{dof} = 0.67$ for $N_{dof}=24$. An attempt to take $n_{max}=3$  and compute $a_5$ gave $a_5=0.008(20)$ with practically
unchanged $a_1$ and $a_3$. Thus $a_5$ is not computable in this case.
In opposite, at lower temperature  $T/T_c=1.20$ we needed ftting function with $n_{max}=3$. We obtained  good fit in this case.

\begin{figure}[htb]
\centering
\includegraphics[width=0.35\textwidth,angle=270]{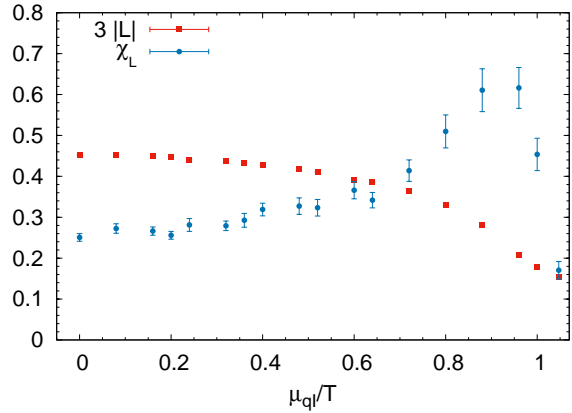}%
\vspace{0cm}
\caption{The absolute value of the Polyakov loop (multiplied by factor 3) and its susceptibility vs. $\mu_{qI}$ at  $T/T_c=1.08$. }
\label{absPL}
\end{figure}

The behavior of  $n_{qI}$  at $T=1.08$ is different from that at higher temperatures discussed above. This temperature is below $T_{RW}$ and at  $\theta=\pi/3$ there is no first order
phase transition, $n_{qI}$ is continuous. Instead there is a crossover to the confinement phase
at about $\theta=0.92(2)$ as is indicated by the Polyakov loop susceptibility, see Fig.~\ref{absPL}.

It is not yet clear how to fit the data over the range of $\mu_{qI}$ covering both deconfining and confining phase.
Our data can be well fitted to various functions which give rise to very different behavior of the number density at real chemical
potential. We need to increase our statistics substantially to reduce the number of functions suitable for fitting of our data.
For this reason in this work we made fit to  function (\ref{eq_fit_polyn}) over the range $[0,0.8]$, i.e. for deconfining phase.
In this case we should consider the fit as a Taylor expansion.
We needed again $n_{max}=3$ to obtain  a good fit. Further increasing of $n_{max}$ produced no reasonable results.

Now we can compare our results for constants $a_i$  with the Taylor expansion results obtained in \cite{Ejiri:2009hq}
(Table IV, lower part) for same
parameters of lattice QCD action as we used in our work. We find that there is an agreement between our values for $a_{1,3}$ and
respective values obtained in \cite{Ejiri:2009hq} within error bars (note that relations between our constants $a_{1,3}$ and
constants $c_{2,4}$ used in \cite{Ejiri:2009hq} are: $c_2=a_1/2, c_4=- a_3/4$).
It should be noted that for $T/T_c=1.35$ and 1.20 where we produced many data points our error bars are
substantially lower than error bars quoted in  \cite{Ejiri:2009hq} (at $T/T_c=1.08$ smaller error bars can be also
achieved after increasing number of data points). Furthermore, we were able to compute coefficient $a_5$ while in
\cite{Ejiri:2009hq} corresponding coefficient $c_6$ was not computed due to complexity of the problem.
This usefullness of simulations at imaginary chemical potential
to compute the Taylor expansion coefficients for pressure was suggested in \cite{Laermann:2013lma}. This was
recently confirmed in \cite{Gunther:2016vcp} where $2+1$ lattice QCD
with physical quark masses and in the continuum limit was studied. We confirm here their observations for completely
different set of parameters and different action of lattice QCD.

Next we compute $Z_n$ using the procedure described in the previous section. Eq.~(\ref{Fourier_2}) now becomes
\beq
Z_n = \frac{  \int_0^{2\pi}\frac{d\theta}{2\pi} \cos(n\theta) e^{-\frac{1}{\cal{N}}\sum_m a_{2m-1}\theta^{2m}/(2m)} }
{  \int_0^{2\pi}\frac{d\theta}{2\pi} e^{-\frac{1}{\cal{N}}\sum_m a_{2m-1}\theta^{2m}/(2m)} }
\label{Fourier_deconf}
\eeq

\begin{figure}[htb]
\centering
\hspace*{0cm}
\includegraphics[width=0.37\textwidth,angle=270]{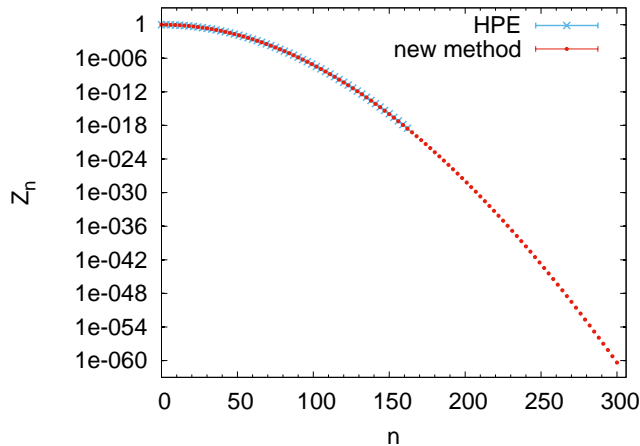}
\vspace{0cm}
\caption{$Z_n$ vs $n$ computed using two methods at $T/T_c=1.35$ }
\label{Zn_deconf_compar}
\end{figure}

We computed these integrals numerically using the multi-precision library \cite{FMlib}.
Results for $Z_n$  are presented in Fig.~\ref{Zn_deconf_compar} for $n$ up to $300$.

As a first check of the results for $Z_n$ we computed $n_{qI}/T^3$ using eq.~(\ref{density2}).
We find that the original data presented in Fig.~\ref{density_deconf} were reproduced nicely.
The deviation for the full interval $[0.0 ; 1.0]$ was less than $0.6 \%$.

As a more important check we compare our results for $Z_n$ with $Z_n$ obtained via hopping parameter expansion which
was described in the previous section. The number of terms in eq.(\ref{eq:Det_expansion_im_muq}) was taken equal to 60.
Results obtained wth HPE method are also presented in Fig.~\ref{Zn_deconf_compar}.
We use only 54 configurations to compute $Z_n$ with HPE. For this reason the statistical error is much higher for this method and
we show only data up to $n=162$ when relative statistical error reaches 50

\begin{figure}[htb]
\centering
\hspace*{0cm}
\includegraphics[width=0.36\textwidth,angle=270]{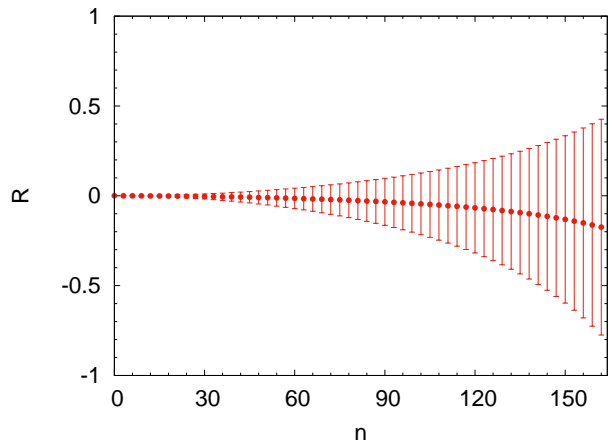}
\vspace{0cm}
\caption{Relative deviation of results for $Z_n$ obtained by new and HPE methods vs $n$ at $T/T_c=1.35$ }
\label{Zn_relative_dev}
\end{figure}

One can see very nice agreement between two results although the values of $Z_n$ change by almost 20 orders of magnitude.
As a more careful check of this agreement we show in Fig.~\ref{Zn_relative_dev} the relative deviation of two results:
\beq
R = \frac{Z_{n,1} - Z_{n,2}}{Z_{n,1}},
\eeq
where $Z_{n,i}, i=1,2$ are for new method and HEP, respectively.
One can see that the relative deviation is compatible with zero for all presented values of $n$. Large statistical errors for large
values of $n$ comes from HPE result and can be reduced when full available statistics is used.

Although both methods we use to compute $Z_n$ are approximate
their systematic errors are of different nature. Thus agreement of results is not a coincidence.
The fact that we obtained correct values for $Z_n$ implies that the fit
(\ref{eq_fit_polyn}) with $n_{max}=2$
is not just a Taylor expansion valid for small values of the chemical potential only but rather a good approximation valid for also for large values of the chemical potential. The range of validity depends on the range of values of index $n$ for which
the agreement between two methods extends.
This statement should be further checked by computation of $Z_n$ via HPE method
for higher $n$ as well as for other temperatures in the deconfinement phase.

We fitted $Z_n$ to function $e^{P(n)}$ where $P(n)$ is a polynomial function
\beq
P(n)= p_2 n^2 + p_4 n^4 + p_6 n^6
\label{asymptotics-1}
\eeq
We found that this function fits the data extremely well:  $\chi^2/N_{dof} =0.028$ for $N_{dof}=97$.
Small $\chi^2/N_{dof}$ might indicate overestimated statistical errors.
Parameter values obtained are as follows: $p_2=-0.00167592(6),\,\, p_4=1.909(4)10^{-9},\,\,  p_6=-5.08(6) 10^{-15} $.
The relative deviation of the fittng function from the data is less then $0.08\%$. When the fit is made over the range $n \in [0,100]$
then prediction for $100 <  n < 300$ has relative deviation from true result less than $1.5\%$. This illustrates the usefullness
of the fitting of the data for $Z_n$ in the deconfinement phase to function $e^{P(n)}$ when the data are known only for the restricted
range of index $n$ values. One can make extrapolation to much higher values of $n$ with rather small error of extrapolation.

\begin{table*}[th]
\begin{tabular}{|c|c|c|c|c|c|c|c|c|} \hline
$T/T_c$  & $f_3$ & $ f_6$& $ a_1$ &$a_3$&$a_5$& $\chi^2/N_{dof}, N_{dof}$  & $ 2c_2$ & $ 24 c_4$ \\  \hline
0.99    &  0.7326(25)&-0.0159(21)& 2.102(5)    &-2.719(17)&   0.453(55) & 0.83, 18 &  2.071(34) & 17.4(47) \\
0.93    & 0.2608(8)  & -          & 0.7824(24) &-1.1736(36) &0.5281(16)& 0.93, 37&  0.713(40) & 2.0(48) \\
0.84    &0.0844(7)   & -          & 0.2532(21) &-0.3798(31)&  0.1709(14)& 0.41, 18&  0.251(35)& 0.0(37) \\ \hline
\end{tabular}
\caption{ Results of fitting data for $n_{qI}/T^3$ in the confinement phase to function (\ref{eq_fit_fourier}). }
\label{table_results_2}
\end{table*}

\begin{figure}[htb]
\centering
\includegraphics[width=0.36\textwidth,angle=270]{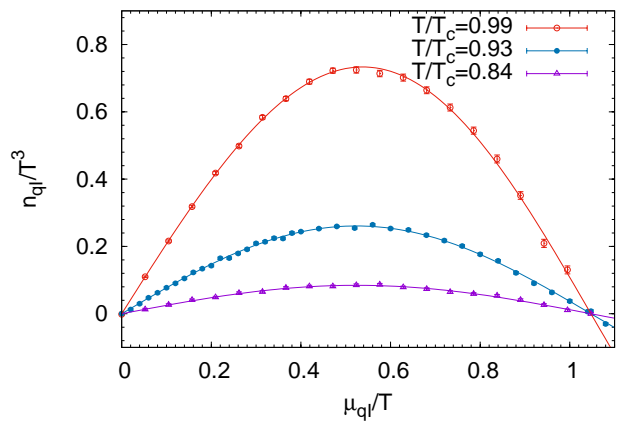}
\vspace{0cm}
\caption{Imaginary density as function of $\theta$ at three temperatures in the confining phase. The curves show fits to function (\ref{eq_fit_fourier}) with $n_{max}=1$ for $T/T_c=0.84, 0.93$ and $ n_{max}=2$ for $T/T_c=0.99$.  }
\label{density_conf}
\end{figure}

Next we come to the confining phase results.
In Fig.~\ref{density_conf} we show $n_{qI}$ for $\theta \in [0,\pi/3]$ together with
fits to eq.~(\ref{eq_fit_fourier}). The fit results are presented in Table~\ref{table_results_2}.
We found good fits with $n_{max}=1$ for $T/T_c=0.84, 0.93$ while for $T/T_c=0.99$ fit with $n_{max}=2$ is necessary.

In Table~\ref{table_results_2} we also show coefficients $a_1, a_3$ and $a_5$ of the Taylor expansion of eq.~(\ref{eq_fit_fourier}) as well as  respectve results from  \cite{Ejiri:2009hq}.
We see again agreement for first two Taylor expansion coefficients within error bars and substantially smaller error bars in our work than in\cite{Ejiri:2009hq}. The third  coefficient was not computed in \cite{Ejiri:2009hq}.
We shall note that dependence of our third Taylor coefficient $a_5$ on the temperature is in qualitative agreement
with results of Refs.~\cite{Allton:2005gk} and \cite{Gunther:2016vcp} where $N_f=2+1$ lattice QCD was studied. In both papers this coefficient ($c_6$ in notations of  ~\cite{Allton:2005gk} and  \cite{Gunther:2016vcp}, $c_6=a_5/6$ ) was found positive slightly below $T_c$, negative slightly above $T_c$ and zero otherwise.

\begin{figure}[htb]
\centering
\hspace*{0cm}
\includegraphics[width=0.36\textwidth,angle=270]{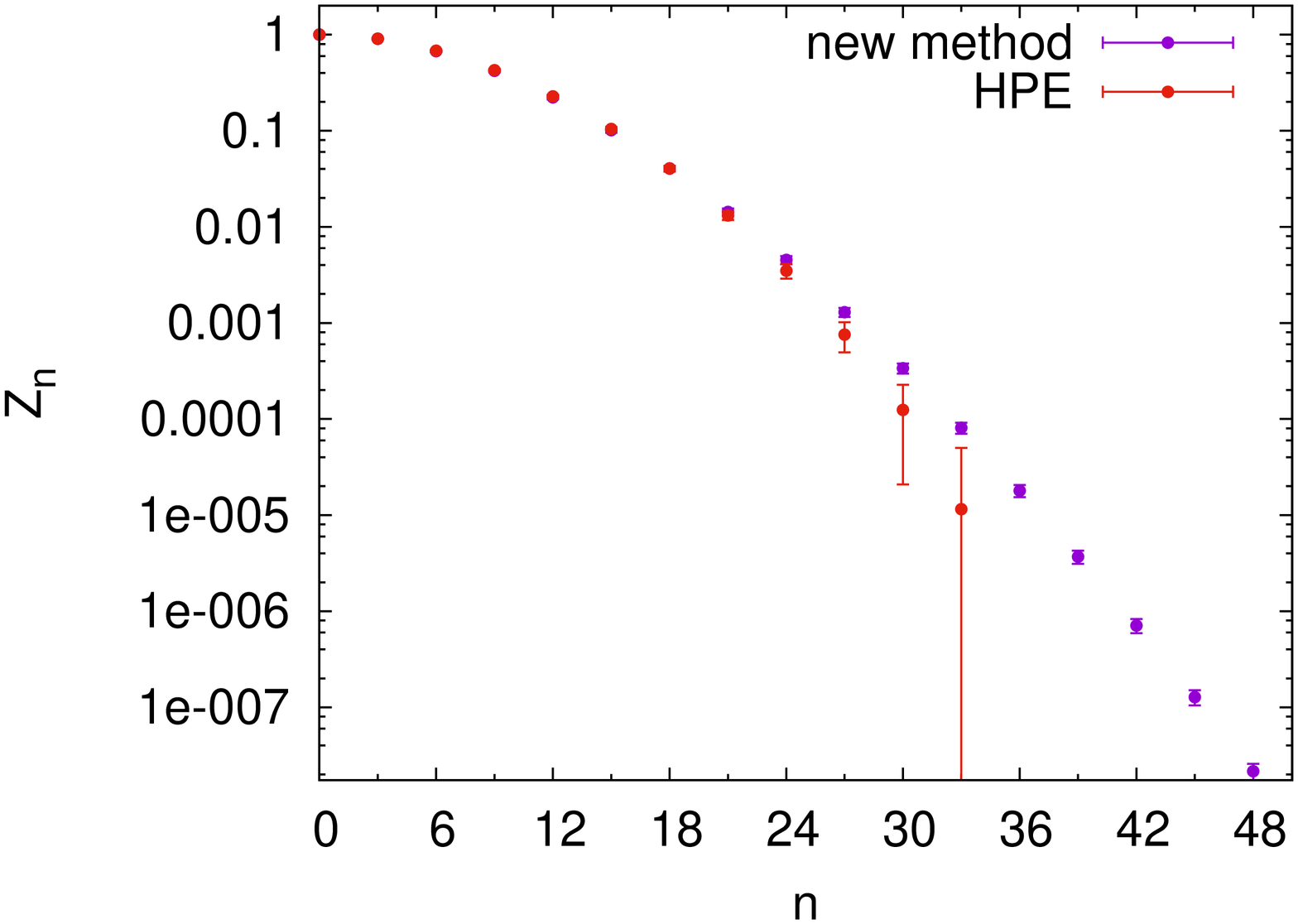}%
\vspace{0cm}
\caption{$Z_n$ vs $n$ computed using two methods at $T/T_c=0.93$ }
\label{Zn_conf_compar}
\end{figure}

Eq.~(\ref{Fourier_2}) to compute $Z_n$ now looks as follows:
\beqa
Z_{3n} &=& \frac{ \int_0^{2\pi}\frac{d\theta}{2\pi} \cos(3n\theta) e^{\sum_{m=1}^{n_{max}} \tilde{f}_{3m} \cos(3m \theta) }}
{\int_0^{2\pi}\frac{d\theta}{2\pi} e^{ \sum_{m=1}^{n_{max}}  \tilde{f}_{3m} \cos(3m \theta)}}
\\
&=& \frac{ \int_0^{6\pi}\frac{dx}{6\pi} \cos(nx) e^{\sum_{m=1}^{n_{max}} \tilde{f}_{3m}  \cos(mx)/  }}
{\int_0^{6\pi}\frac{dx}{6\pi}  e^{\sum_{m=1}^{n_{max}} \tilde{f}_{3m} \cos(mx)/}}
\label{Fourier_conf}
\eeqa
where $\tilde{f}_{3n}=\frac{N_s^3}{N_t^3} \frac{f_3}{3n}$.
In case $n_{max}=1$ this can be expressed as
\beq
Z_{3n} =  \frac{I_n(\tilde{f}_3)}{I_0(\tilde{f}_3)}\, ,
\eeq
where $I_n(x)$ is the modified Bessel function of the 1st kind.
Results for $Z_n$ at $T/T_c=0.93$ are presented in Fig.~\ref{Zn_conf_compar}.

\begin{figure}[htb]
\centering
\hspace*{0cm}
\includegraphics[width=0.36\textwidth,angle=270]{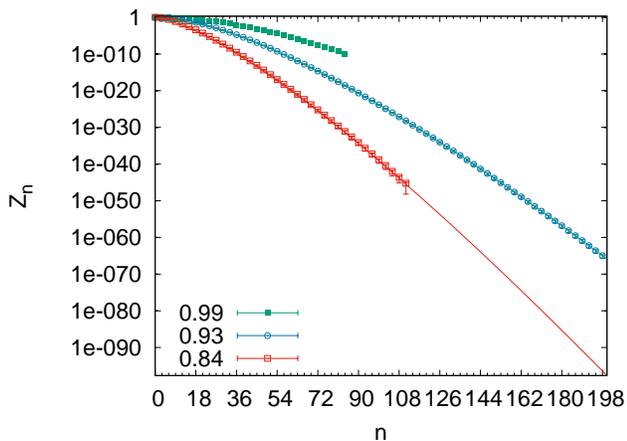}%
\vspace{0cm}
\caption{$Z_n$ vs $n$ for three temperatures in the confinement phase. The curves show results of fitting to asymptotic behavior
for $T/T_c=0.93$ and 0.84. }
\label{Zn_conf_all}
\end{figure}

Again, we first check if computed $Z_n$ reproduce data for $n_{qI}/T^3$ and find nice agreement
between data and values of $n_{qI}/T^3$ computed via eq.~(\ref{density2}).  The deviation
for the full interval [0.0; $\pi/3$] was less than $0.3\%$.

Next we compare with hopping parameter expansion, respective results are also presented in Fig.~\ref{Zn_conf_compar}.
We used full statistics (1800 configurations at $\mu_{qI}=0$) and $W_n$ up to $n=15$ in eq.~(\ref{eq:Det_expansion_W_n}) for this computation.
One can see agreement between two results up to $n=21$.
We believe that the disagreement for higher $n$ is explained by inaccuracy in computation of $Z_n$ computed by HPE.
The statistical errors for $Z_n$ grow very fast with $n$. It is necessary to improve the HPE method accuracy before the
conclusion about agreement at large $n$ can be made.

Still our result indicates that at $T/T_c=0.93$ the fit function (\ref{eq_fit_fourier}) provides correct values of
$Z_n$ and thus its analytical continuation should be valid up to  values of $\mu_q/T$ beyond Taylor expansion validity
range. Precise determination
of the range of validity of this analytical continuation will be made in future after getting more precise
results for HPE method.  In Fig.~\ref{Zn_conf_all} we show results for $Z_n$ for all three temperatures below $T_c$.
We stop to show data in this figure when the relative statistical error reaches 100\%.

Let us note that one can derive a recursion relation for $Z_n$ when $n_{qI}$ is presented by
function (\ref{eq_fit_fourier}) with finite $n_{max}$.
For derivation see Appendix A. In particular for $n_{max}=1$ the recursion is just a recursion for $I_n$
which is of the form
\beq
  \tilde{f_3}( Z_{3(n-1)} - Z_{3(n+1)} ) =  2n Z_{3n}.
\label{recursion_1}
\eeq

Also one can get asymptotics for $Z_n$ at large $n$. For $n_{max}=1$ it is
\beq
Z_{3n} = B \frac{(\tilde{f}_3/2)^n}{n !},
\label{asymptotics-2}
\eeq
where $B$ is some  constant. This asymptotics is shown in Fig.~\ref{Zn_conf_all} for
$T/T_c=0.84$ and 0.93 with constant $B$ obtained by fitting over the range $400 < n < 600$:
$B=0.02813((1)$ for $T/T_c=0.93$ and  $B=0.219(2)$ for $T/T_c=0.84$ .

For  $n_{max}=N$ it is different:
 \beq
Z_{3n} = B\frac{(\tilde{f}_{3N})^{n/N}}{\Gamma(n/N+1)},
\label{asymptotics-3}
\eeq
see  Appendix A for derivation. From this asymptotics it follows in particular that the coefficient
$f_{3N}$ has to be positive, otherwise the condition of positivity of $Z_{3n}$ will not hold.
Our current fitting function for $T/T_c=0.99$ which has $f_{3N} \equiv f_6<0$ does not satisfy
this requirement. We need to improve statistics for this temperature to obtain the coefficcient
for the next harmonics in eq.~(\ref{eq_fit_fourier}). Evidently with present fitting function we cannot
go to large values of $\mu_q$ for this temperature.

\begin{figure}[htb]
\centering
\hspace*{0cm}
\includegraphics[width=0.36\textwidth,angle=270]{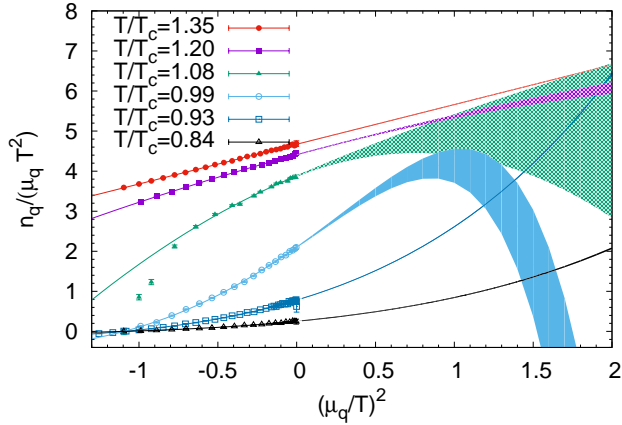}%
\vspace{0cm}
\caption{Analytical continuation for the number density vs. $\mu_q^2$ for all temperature values.
The curves show respective fits. The width of the curves  shows the statistical error of extrapolation to $\mu_q^2 > 0$.}
\label{dens_anal_cont}
\end{figure}

At the end of this section we show in Fig.~\ref{dens_anal_cont} ratio $\frac{n_q}{\mu_q T^2}$ as function of $\mu_q^2$ for negative and positive
values. This way of presentation, borrowed from \cite{Gunther:2016vcp} allows to show in one plot the simulation
results obtained at  $\mu_q^2 < 0$
and analytical continuation of our fitting functions to  $\mu_q^2 > 0$.
One can see that analytical continuation has reasonable statistical errors up to large
values of $(\mu_q/T)^2$ for two highest temperatures and two lowest temperatures. For temperatures $T/T_c=0.99, 1.08$ we need to
improve statistics.

Let us note that another possibility to check the range of validity of the new method suggested in this paper is to make simulations in a model without sign problem like QC$_2$D~\cite{Makiyama:2015uwa, Braguta:2016cpw}. We are planning to do such checks in future.

\section{Conclusions}
\label{conclusions}

We have presented new method to compute the canonical partition functions $Z_n$.
It is based on fitting of the imaginary number density for all values of imaginary
chemical potential to the theoretically motivated
fitting functions:  polynomial fit (\ref{eq_fit_polyn}) in the deconfinement
phase for $T$ above $T_{RW}$ and Fourier-type fit (\ref{eq_fit_fourier}) in the confinement phase.
The proper fit for temperatures between $T_c$ and $T_{RW}$ has not been found
in this work and is a subject of future study. For this temperature range we used the polynomial fit for
restricted range of $\mu_{qI}$.

Using fit results we compute canonical partition functions $Z_n \equiv \frac{Z_C(n,T,V)}{Z_C(0,T,V)}$ at 5 values of $T/T_c$
(all temperatures apart from $T/T_c=1.08$)
via Fourier transformation (\ref{Fourier_2}). It was necessary to use the multi-precision library \cite{FMlib} to compute
$Z_n$ which change over many orders of magnitude. For all temperatures  we have checked that precision
of computation of $Z_n$ was high enough to reproduce the imaginary number density $n_{qI}$  via eq.~(\ref{density2}).

At temperatures  $T/T_c=1.35$ and 0.93 we compared our results for $Z_n$ with $Z_n$ computed by hopping parameter expansion.
We found that the new method works in both confinement and deconfinement phases: two sets of $Z_n$ computed by
completely independent methods agree well, see Figs.~\ref{Zn_deconf_compar}, \ref{Zn_conf_compar}.
This means that the fitting functions used in this work
are proper approximations for the imaginary number density in the full range of $\mu_{qI}$ values.
Furthermore, this means that the analytical continuation to the real chemical potential can in principle be done
beyond the Taylor expansion
validity range since this analytical continuation coincides with $n_q$ computed with the help of correctly determined
$Z_n$ via eq.~(\ref{density_re}). Thus the new method allows to compute the number density $n_q$ beyond Taylor expansion.
The range of validity of new method is implicitly determined by the number of correctly computed $Z_n$. This number can be
increased by increasing  the quality of approximation of imaginary number density. This can  be achieved
when more terms in eqs.~(\ref{eq_fit_polyn}), (\ref{eq_fit_fourier}) are determined via fitting procedure or via
direct numerical computation of the integral (\ref{integration_1}).

The agreement of the new method and HPE method is especially remarkable in the deconfining phase, see Figs.~\ref{Zn_deconf_compar} and
\ref{Zn_relative_dev}.
The deconfinement region is being explored extensively by ALICE experiments
at LHC \cite{Aamodt:2008zz}. Note that our new method is not limited to the
heavy quark mass values like HPE, nor small $\mu$ values like Taylor expansion.
Once we calculate $Z_n$ using new method, we can calculate any thermodynamical
quantities, pressure,  number density and its higher moments.
Thus the new method can provide very reliable theoretical basis for LHC results.
We plan to calculate these quantities with much smaller quark mass in order to
give first-principle theoretical results for comparison with LHC data.

We believe that the new method will
also help to determine the transition line in the temperature - chemical potential plane. Respective results will be presented
in a forthcoming publication after data with higher statistics will be accumulated and results for lower quark masses
will be obtained.

Using our results for the number density $n_{qI}$ we computed the Taylor expansion coefficients for the number density from which respective coefficients for the pressure are
easily restored. We found good agreement with earlier results obtained in \cite{Ejiri:2009hq} via direct computation of these coefficients.
Moreover we found that our error bars for these coefficients are in general substantially smaller than  error bars
quoted in \cite{Ejiri:2009hq}. Thus we confirmed analogous  observation made in \cite{Gunther:2016vcp}. Our estimation for 6-th order
Taylor coefficients $c_6$ which were not computed in  \cite{Ejiri:2009hq} are in good qualitative agreement with results
of Refs.~\cite{Allton:2005gk} and \cite{Gunther:2016vcp}.

We found that obtained at $T/T_c=1.35$  $Z_n$ values are nicely described by the exponential behavior with polynomial
(\ref{asymptotics-1}) in the exponent. We checked that this fit works over the range of $n$ up to 300 which corresponds to quark density $n_q/T^3 \approx 5$.
For $T < T_c$ we  obtained asymptotics of $Z_n$ at large $n$ which
indicates slower decreasing of $Z_n$ with $n$ than in the deconfining phase, see eq.~(\ref{asymptotics-3}). Still this decreasing
is fast enough to provide convergence of the infinite  sums in eqs.~(\ref{ZG}) and (\ref{density_re}).

\acknowledgments{The authors are grateful to Ph. de Forcrand and M. Hanada for useful discussions. This work was completed due to support by
RSF grant under contract 15-12-20008. Computer simulations were performed on the FEFU GPU cluster Vostok-1 and  MSU 'Lomonosov' supercomputer. }

\vspace{.5cm}
\centerline{
\appendix{\bf Appendix A}
}
\vspace{.5cm}
In this Appendix we derive a recursion relation for $Z_{3n}$ in the confinement phase.
Let us introduce notations $3\theta=x$, $\tilde{f}_n=f_n/(2C)$.

For the case of $n_{max}>1$ in eq.(\ref{eq_fit_fourier}) it is possible to derive a recursion relations
similar to relation (\ref{recursion_1}).
Let us show this for  $n_{max}=2$.
We have
\beq
\tilde{f}_3 \sin (x) +  \tilde{f}_6 \sin (2x) = \frac{\sum_n n Z_{3n} \sin (n x)} {1 + 2  \sum_n Z_{3n} \cos( n x)}.
\eeq
Then
\beqa
&& ( \tilde{f}_3 \sin (x) +  \tilde{f}_6 \sin (2x) ) ( 1 + 2  \sum_n Z_{3n} \cos( n x) )  \nonumber \\
&=& \sum_n n Z_{3n} \sin (n x). \nonumber
\eeqa

Computing Fourier modes on both sides we get:
\beqa
&& \tilde{f}_3 \int_{-\pi}^{\pi} dx \sin (x) \left( 1 + 2  \sum_n  Z_{3n} \cos( n x) \right) \sin(m x)
\nonumber \\
&+& \tilde{f}_6 \int_{-\pi}^{\pi} dx \sin (2x) \left( 1 + 2  \sum_n  Z_{3n} \cos( n x) \right) \sin(m x) \nonumber \\
 &=&  \int_{-\pi}^{\pi}  dx  \sum_n n Z_{3n} \sin (n x) \sin(m x)
\eeqa

\beq
 \tilde{f}_3 ( Z_{3(m-1)} - Z_{3(m+1)} ) + \tilde{f}_6 ( Z_{3(m-2)} - Z_{3(m+2)} )  =   m Z_{3m}
\eeq
or
\beqa
Z_{3(m+2)} &=&   Z_{3(m-2)} - \frac{m}{\tilde{f}_6} Z_{3m} + \nonumber \\
&& \frac{\tilde{f}_3}{\tilde{f}_6} ( Z_{3(m-1)} - Z_{3(m+1)} )
\eeqa
We need $f_3$, $f_6$, $Z_3$, $Z_6$ to compute all $Z_{3m}, m>2$.

The asymptotical behavior in this case is
\beq
Z_{3n} = B\frac{(\tilde{f}_6)^{n/2}}{\Gamma(n/2+1)} \\
\label{eq:asympt2}
\eeq
It is easy to get the recursion relation and asymptotics for $n_{max}=N$. The recursion
relation is
\beqa
Z_{3(n+N)} &=&   Z_{3(n-N)} - \frac{n}{\tilde{f}_{3N}} Z_{3n} + \nonumber \\
&& \sum_{m=1}^{N-1}\frac{\tilde{f}_{3(n-m)}}{\tilde{f}_{3N}} ( Z_{3(n-m)} - Z_{3(n+m)} )
\eeqa
and the asymptotics
 \beq
Z_{3n} = B\frac{(\tilde{f}_{3N})^{n/N}}{\Gamma(n/N+1)},
\eeq
Some conclusions might be drawn from this expression. The asymptotics is determined by the highest mode.
Thus $f_{3N}$ has to be positive. Decreasing of $Z_n$ becomes weaker with increasing $N$.

The numerical data for $n_{qI}(\theta)$ indicate that in the confinement phase the number of modes $N$ necessary
to describe the data is finite. Then the above considerations apply. And we can make a statement that
the radius of convergence is infinite. In the deconfinement phase at temperatures
$T>T_{RW}$ where first order Roberge - Weiss transition takes place $N$ is definitely infinite. In the range of temperature
$T_c < T < T_{RW}$ the situation is unclear.


\begin{thebibliography}{26}
\expandafter\ifx\csname natexlab\endcsname\relax\def\natexlab#1{#1}\fi
\expandafter\ifx\csname bibnamefont\endcsname\relax
  \def\bibnamefont#1{#1}\fi
\expandafter\ifx\csname bibfnamefont\endcsname\relax
  \def\bibfnamefont#1{#1}\fi
\expandafter\ifx\csname citenamefont\endcsname\relax
  \def\citenamefont#1{#1}\fi
\expandafter\ifx\csname url\endcsname\relax
  \def\url#1{\texttt{#1}}\fi
\expandafter\ifx\csname urlprefix\endcsname\relax\def\urlprefix{URL }\fi
\providecommand{\bibinfo}[2]{#2}
\providecommand{\eprint}[2][]{\url{#2}}

\bibitem[{\citenamefont{Adams et~al.}(2005)}]{Adams:2005dq}
\bibinfo{author}{\bibfnamefont{J.}~\bibnamefont{Adams}} \bibnamefont{et~al.}
  (\bibinfo{collaboration}{STAR}), \bibinfo{journal}{Nucl. Phys.}
  \textbf{\bibinfo{volume}{A757}}, \bibinfo{pages}{102} (\bibinfo{year}{2005}),
  \eprint{nucl-ex/0501009}.

\bibitem[{\citenamefont{Aamodt et~al.}(2008)}]{Aamodt:2008zz}
\bibinfo{author}{\bibfnamefont{K.}~\bibnamefont{Aamodt}} \bibnamefont{et~al.}
  (\bibinfo{collaboration}{ALICE}), \bibinfo{journal}{JINST}
  \textbf{\bibinfo{volume}{3}}, \bibinfo{pages}{S08002} (\bibinfo{year}{2008}).

\bibitem[{\citenamefont{Muroya et~al.}(2003)\citenamefont{Muroya, Nakamura,
  Nonaka, and Takaishi}}]{Muroya:2003qs}
\bibinfo{author}{\bibfnamefont{S.}~\bibnamefont{Muroya}},
  \bibinfo{author}{\bibfnamefont{A.}~\bibnamefont{Nakamura}},
  \bibinfo{author}{\bibfnamefont{C.}~\bibnamefont{Nonaka}}, \bibnamefont{and}
  \bibinfo{author}{\bibfnamefont{T.}~\bibnamefont{Takaishi}},
  \bibinfo{journal}{Prog. Theor. Phys.} \textbf{\bibinfo{volume}{110}},
  \bibinfo{pages}{615} (\bibinfo{year}{2003}), \eprint{hep-lat/0306031}.

\bibitem[{\citenamefont{Philipsen}(2006)}]{Philipsen:2005mj}
\bibinfo{author}{\bibfnamefont{O.}~\bibnamefont{Philipsen}},
  \bibinfo{journal}{PoS} \textbf{\bibinfo{volume}{LAT2005}},
  \bibinfo{pages}{016} (\bibinfo{year}{2006}),
  \bibinfo{note}{[PoSJHW2005,012(2006)]}, \eprint{hep-lat/0510077}.

\bibitem[{\citenamefont{de~Forcrand}(2009)}]{deForcrand:2010ys}
\bibinfo{author}{\bibfnamefont{P.}~\bibnamefont{de~Forcrand}},
  \bibinfo{journal}{PoS} \textbf{\bibinfo{volume}{LAT2009}},
  \bibinfo{pages}{010} (\bibinfo{year}{2009}), \eprint{1005.0539}.

\bibitem[{\citenamefont{Nagata and Nakamura}(2012)}]{Nagata:2012pc}
\bibinfo{author}{\bibfnamefont{K.}~\bibnamefont{Nagata}} \bibnamefont{and}
  \bibinfo{author}{\bibfnamefont{A.}~\bibnamefont{Nakamura}},
  \bibinfo{journal}{JHEP} \textbf{\bibinfo{volume}{04}}, \bibinfo{pages}{092}
  (\bibinfo{year}{2012}), \eprint{1201.2765}.

\bibitem[{\citenamefont{de~Forcrand and Kratochvila}(2006)}]{deForcrand:2006ec}
\bibinfo{author}{\bibfnamefont{P.}~\bibnamefont{de~Forcrand}} \bibnamefont{and}
  \bibinfo{author}{\bibfnamefont{S.}~\bibnamefont{Kratochvila}},
  \bibinfo{journal}{Nucl. Phys. Proc. Suppl.} \textbf{\bibinfo{volume}{153}},
  \bibinfo{pages}{62} (\bibinfo{year}{2006}), \bibinfo{note}{[,62(2006)]},
  \eprint{hep-lat/0602024}.

\bibitem[{\citenamefont{Ejiri}(2008)}]{Ejiri:2008xt}
\bibinfo{author}{\bibfnamefont{S.}~\bibnamefont{Ejiri}},
  \bibinfo{journal}{Phys. Rev.} \textbf{\bibinfo{volume}{D78}},
  \bibinfo{pages}{074507} (\bibinfo{year}{2008}), \eprint{0804.3227}.

\bibitem[{\citenamefont{Li et~al.}(2010)\citenamefont{Li, Alexandru, Liu, and
  Meng}}]{Li:2010qf}
\bibinfo{author}{\bibfnamefont{A.}~\bibnamefont{Li}},
  \bibinfo{author}{\bibfnamefont{A.}~\bibnamefont{Alexandru}},
  \bibinfo{author}{\bibfnamefont{K.-F.} \bibnamefont{Liu}}, \bibnamefont{and}
  \bibinfo{author}{\bibfnamefont{X.}~\bibnamefont{Meng}},
  \bibinfo{journal}{Phys. Rev.} \textbf{\bibinfo{volume}{D82}},
  \bibinfo{pages}{054502} (\bibinfo{year}{2010}), \eprint{1005.4158}.

\bibitem[{\citenamefont{Li et~al.}(2011)\citenamefont{Li, Alexandru, and
  Liu}}]{Li:2011ee}
\bibinfo{author}{\bibfnamefont{A.}~\bibnamefont{Li}},
  \bibinfo{author}{\bibfnamefont{A.}~\bibnamefont{Alexandru}},
  \bibnamefont{and} \bibinfo{author}{\bibfnamefont{K.-F.} \bibnamefont{Liu}},
  \bibinfo{journal}{Phys. Rev.} \textbf{\bibinfo{volume}{D84}},
  \bibinfo{pages}{071503} (\bibinfo{year}{2011}), \eprint{1103.3045}.

\bibitem[{\citenamefont{Danzer and Gattringer}(2012)}]{Danzer:2012vw}
\bibinfo{author}{\bibfnamefont{J.}~\bibnamefont{Danzer}} \bibnamefont{and}
  \bibinfo{author}{\bibfnamefont{C.}~\bibnamefont{Gattringer}},
  \bibinfo{journal}{Phys. Rev.} \textbf{\bibinfo{volume}{D86}},
  \bibinfo{pages}{014502} (\bibinfo{year}{2012}), \eprint{1204.1020}.

\bibitem[{\citenamefont{Gattringer and Schadler}(2015)}]{Gattringer:2014hra}
\bibinfo{author}{\bibfnamefont{C.}~\bibnamefont{Gattringer}} \bibnamefont{and}
  \bibinfo{author}{\bibfnamefont{H.-P.} \bibnamefont{Schadler}},
  \bibinfo{journal}{Phys. Rev.} \textbf{\bibinfo{volume}{D91}},
  \bibinfo{pages}{074511} (\bibinfo{year}{2015}), \eprint{1411.5133}.

\bibitem[{\citenamefont{Fukuda et~al.}(2016)\citenamefont{Fukuda, Nakamura, and
  Oka}}]{Fukuda:2015mva}
\bibinfo{author}{\bibfnamefont{R.}~\bibnamefont{Fukuda}},
  \bibinfo{author}{\bibfnamefont{A.}~\bibnamefont{Nakamura}}, \bibnamefont{and}
  \bibinfo{author}{\bibfnamefont{S.}~\bibnamefont{Oka}},
  \bibinfo{journal}{Phys. Rev.} \textbf{\bibinfo{volume}{D93}},
  \bibinfo{pages}{094508} (\bibinfo{year}{2016}), \eprint{1504.06351}.

\bibitem[{\citenamefont{Nakamura et~al.}(2016)\citenamefont{Nakamura, Oka, and
  Taniguchi}}]{Nakamura:2015jra}
\bibinfo{author}{\bibfnamefont{A.}~\bibnamefont{Nakamura}},
  \bibinfo{author}{\bibfnamefont{S.}~\bibnamefont{Oka}}, \bibnamefont{and}
  \bibinfo{author}{\bibfnamefont{Y.}~\bibnamefont{Taniguchi}},
  \bibinfo{journal}{JHEP} \textbf{\bibinfo{volume}{02}}, \bibinfo{pages}{054}
  (\bibinfo{year}{2016}), \eprint{1504.04471}.

\bibitem[{\citenamefont{Hasenfratz and Toussaint}(1992)}]{Hasenfratz:1991ax}
\bibinfo{author}{\bibfnamefont{A.}~\bibnamefont{Hasenfratz}} \bibnamefont{and}
  \bibinfo{author}{\bibfnamefont{D.}~\bibnamefont{Toussaint}},
  \bibinfo{journal}{Nucl. Phys.} \textbf{\bibinfo{volume}{B371}},
  \bibinfo{pages}{539} (\bibinfo{year}{1992}).

\bibitem[{\citenamefont{Roberge and Weiss}(1986)}]{Roberge:1986mm}
\bibinfo{author}{\bibfnamefont{A.}~\bibnamefont{Roberge}} \bibnamefont{and}
  \bibinfo{author}{\bibfnamefont{N.}~\bibnamefont{Weiss}},
  \bibinfo{journal}{Nucl. Phys.} \textbf{\bibinfo{volume}{B275}},
  \bibinfo{pages}{734} (\bibinfo{year}{1986}).

\bibitem[{\citenamefont{Bonati et~al.}(2014)\citenamefont{Bonati, de~Forcrand,
  D'Elia, Philipsen, and Sanfilippo}}]{Bonati:2014kpa}
\bibinfo{author}{\bibfnamefont{C.}~\bibnamefont{Bonati}},
  \bibinfo{author}{\bibfnamefont{P.}~\bibnamefont{de~Forcrand}},
  \bibinfo{author}{\bibfnamefont{M.}~\bibnamefont{D'Elia}},
  \bibinfo{author}{\bibfnamefont{O.}~\bibnamefont{Philipsen}},
  \bibnamefont{and}
  \bibinfo{author}{\bibfnamefont{F.}~\bibnamefont{Sanfilippo}},
  \bibinfo{journal}{Phys. Rev.} \textbf{\bibinfo{volume}{D90}},
  \bibinfo{pages}{074030} (\bibinfo{year}{2014}), \eprint{1408.5086}.

\bibitem[{\citenamefont{Takahashi et~al.}(2015)\citenamefont{Takahashi, Kouno,
  and Yahiro}}]{Takahashi:2014rta}
\bibinfo{author}{\bibfnamefont{J.}~\bibnamefont{Takahashi}},
  \bibinfo{author}{\bibfnamefont{H.}~\bibnamefont{Kouno}}, \bibnamefont{and}
  \bibinfo{author}{\bibfnamefont{M.}~\bibnamefont{Yahiro}},
  \bibinfo{journal}{Phys. Rev.} \textbf{\bibinfo{volume}{D91}},
  \bibinfo{pages}{014501} (\bibinfo{year}{2015}), \eprint{1410.7518}.

\bibitem[{\citenamefont{Gunther et~al.}(2016)\citenamefont{Gunther, Bellwied,
  Borsanyi, Fodor, Katz, Pasztor, and Ratti}}]{Gunther:2016vcp}
\bibinfo{author}{\bibfnamefont{J.}~\bibnamefont{Gunther}},
  \bibinfo{author}{\bibfnamefont{R.}~\bibnamefont{Bellwied}},
  \bibinfo{author}{\bibfnamefont{S.}~\bibnamefont{Borsanyi}},
  \bibinfo{author}{\bibfnamefont{Z.}~\bibnamefont{Fodor}},
  \bibinfo{author}{\bibfnamefont{S.~D.} \bibnamefont{Katz}},
  \bibinfo{author}{\bibfnamefont{A.}~\bibnamefont{Pasztor}}, \bibnamefont{and}
  \bibinfo{author}{\bibfnamefont{C.}~\bibnamefont{Ratti}}
  (\bibinfo{year}{2016}), \eprint{1607.02493}.

\bibitem[{\citenamefont{Karsch et~al.}(2003)\citenamefont{Karsch, Redlich, and
  Tawfik}}]{Karsch:2003zq}
\bibinfo{author}{\bibfnamefont{F.}~\bibnamefont{Karsch}},
  \bibinfo{author}{\bibfnamefont{K.}~\bibnamefont{Redlich}}, \bibnamefont{and}
  \bibinfo{author}{\bibfnamefont{A.}~\bibnamefont{Tawfik}},
  \bibinfo{journal}{Phys. Lett.} \textbf{\bibinfo{volume}{B571}},
  \bibinfo{pages}{67} (\bibinfo{year}{2003}), \eprint{hep-ph/0306208}.

\bibitem[{\citenamefont{Ejiri et~al.}(2010)\citenamefont{Ejiri, Maezawa, Ukita,
  Aoki, Hatsuda, Ishii, Kanaya, and Umeda}}]{Ejiri:2009hq}
\bibinfo{author}{\bibfnamefont{S.}~\bibnamefont{Ejiri}},
  \bibinfo{author}{\bibfnamefont{Y.}~\bibnamefont{Maezawa}},
  \bibinfo{author}{\bibfnamefont{N.}~\bibnamefont{Ukita}},
  \bibinfo{author}{\bibfnamefont{S.}~\bibnamefont{Aoki}},
  \bibinfo{author}{\bibfnamefont{T.}~\bibnamefont{Hatsuda}},
  \bibinfo{author}{\bibfnamefont{N.}~\bibnamefont{Ishii}},
  \bibinfo{author}{\bibfnamefont{K.}~\bibnamefont{Kanaya}}, \bibnamefont{and}
  \bibinfo{author}{\bibfnamefont{T.}~\bibnamefont{Umeda}}
  (\bibinfo{collaboration}{WHOT-QCD}), \bibinfo{journal}{Phys. Rev.}
  \textbf{\bibinfo{volume}{D82}}, \bibinfo{pages}{014508}
  (\bibinfo{year}{2010}), \eprint{0909.2121}.

\bibitem[{\citenamefont{Laermann et~al.}(2013)\citenamefont{Laermann, Meyer,
  and Lombardo}}]{Laermann:2013lma}
\bibinfo{author}{\bibfnamefont{E.}~\bibnamefont{Laermann}},
  \bibinfo{author}{\bibfnamefont{F.}~\bibnamefont{Meyer}}, \bibnamefont{and}
  \bibinfo{author}{\bibfnamefont{M.~P.} \bibnamefont{Lombardo}},
  \bibinfo{journal}{J. Phys. Conf. Ser.} \textbf{\bibinfo{volume}{432}},
  \bibinfo{pages}{012016} (\bibinfo{year}{2013}), \eprint{1304.3247}.

\bibitem[{\citenamefont{Smith}(2016)}]{FMlib}
\bibinfo{author}{\bibfnamefont{D.~M.} \bibnamefont{Smith}},
  \bibinfo{journal}{\url{http://myweb.lmu.edu/dmsmith/fmlib.html}}
  (\bibinfo{year}{2016}).

\bibitem[{\citenamefont{Allton et~al.}(2005)\citenamefont{Allton, Doring,
  Ejiri, Hands, Kaczmarek, Karsch, Laermann, and Redlich}}]{Allton:2005gk}
\bibinfo{author}{\bibfnamefont{C.~R.} \bibnamefont{Allton}},
  \bibinfo{author}{\bibfnamefont{M.}~\bibnamefont{Doring}},
  \bibinfo{author}{\bibfnamefont{S.}~\bibnamefont{Ejiri}},
  \bibinfo{author}{\bibfnamefont{S.~J.} \bibnamefont{Hands}},
  \bibinfo{author}{\bibfnamefont{O.}~\bibnamefont{Kaczmarek}},
  \bibinfo{author}{\bibfnamefont{F.}~\bibnamefont{Karsch}},
  \bibinfo{author}{\bibfnamefont{E.}~\bibnamefont{Laermann}}, \bibnamefont{and}
  \bibinfo{author}{\bibfnamefont{K.}~\bibnamefont{Redlich}},
  \bibinfo{journal}{Phys. Rev.} \textbf{\bibinfo{volume}{D71}},
  \bibinfo{pages}{054508} (\bibinfo{year}{2005}), \eprint{hep-lat/0501030}.

\bibitem[{\citenamefont{Makiyama et~al.}(2016)\citenamefont{Makiyama, Sakai,
  Saito, Ishii, Takahashi, Kashiwa, Kouno, Nakamura, and
  Yahiro}}]{Makiyama:2015uwa}
\bibinfo{author}{\bibfnamefont{T.}~\bibnamefont{Makiyama}},
  \bibinfo{author}{\bibfnamefont{Y.}~\bibnamefont{Sakai}},
  \bibinfo{author}{\bibfnamefont{T.}~\bibnamefont{Saito}},
  \bibinfo{author}{\bibfnamefont{M.}~\bibnamefont{Ishii}},
  \bibinfo{author}{\bibfnamefont{J.}~\bibnamefont{Takahashi}},
  \bibinfo{author}{\bibfnamefont{K.}~\bibnamefont{Kashiwa}},
  \bibinfo{author}{\bibfnamefont{H.}~\bibnamefont{Kouno}},
  \bibinfo{author}{\bibfnamefont{A.}~\bibnamefont{Nakamura}}, \bibnamefont{and}
  \bibinfo{author}{\bibfnamefont{M.}~\bibnamefont{Yahiro}},
  \bibinfo{journal}{Phys. Rev.} \textbf{\bibinfo{volume}{D93}},
  \bibinfo{pages}{014505} (\bibinfo{year}{2016}), \eprint{1502.06191}.

\bibitem[{\citenamefont{Braguta et~al.}(2016)\citenamefont{Braguta, Ilgenfritz,
  Kotov, Molochkov, and Nikolaev}}]{Braguta:2016cpw}
\bibinfo{author}{\bibfnamefont{V.~V.} \bibnamefont{Braguta}},
  \bibinfo{author}{\bibfnamefont{E.~M.} \bibnamefont{Ilgenfritz}},
  \bibinfo{author}{\bibfnamefont{A.~{\relax Yu}.} \bibnamefont{Kotov}},
  \bibinfo{author}{\bibfnamefont{A.~V.} \bibnamefont{Molochkov}},
  \bibnamefont{and} \bibinfo{author}{\bibfnamefont{A.~A.}
  \bibnamefont{Nikolaev}} (\bibinfo{year}{2016}), \eprint{1605.04090}.

\end{thebibliography}

\end{document}